\documentclass[11pt]{article}
\usepackage{graphicx}
\usepackage[T1]{fontenc}
\usepackage[utf8]{inputenc}
\usepackage{authblk}
\begin{document}

\title{\emph{C. elegans} in Complex Media}%
\author[1]{X. N. Shen}%
\author[1]{G. Juarez}
\author[1]{P. E. Arratia \thanks{parratia@seas.upenn.edu}}
\affil[1]{Department of Mechanical Engineering and Applied Mechanics, University of Pennsylvania, Philadelphia, PA 19104, USA}

\date{\today}%

\maketitle

\begin{abstract}
We experimentally studied the locomotion of the nematode \emph{C. elegans} in both fluidic and granular media. In this fluid dynamics video, we show the motility gaits of the nematode in these two environments. The motility of the nematode \emph{C. elegans} is investigated using particle tracking methods. Experimental results show that different transport patterns emerge from the fluidic and granular media during the nematode locomotion.
\end{abstract}
\section{Introduction}
Many microorganisms have dwelled and evolved in complex media, including water, soil and mud. The material properties played an important role in the motilities of the microorganisms. As the nematode moves through the soil, the nematode exhibits curly `S' body shapes which are commonly referred as ``crawling''. On the other hand, the nematode bends less yet faster in simple fluids like water (or M9 buffer solution), which are usually termed as ``swimming''. In this video, we studied the nematode motility gaits in fluidic and granular media and investigated the particle transport patterns emerged from these two environments.

The fluidic environment consists of the M9 buffer solution ( the fluid viscosity $\mu \sim 1.0$ Pa$\cdot$s)~\cite{JosuePF}. Granular environments are prepared with monodisperse glass beads with the diameter of $d$ = 60 $\pm$ 3 $\mu$m and polydisperse glass beads with the diameter $d$ = 52 $\pm$ 10 $\mu$m. In the video shown, the packing area density of the granular media $\phi$ is approximately 0.55 for both monodisperse and polydisperse beads. The motility gaits in these two environments have been imaged and recorded through an inverted microscope via the high-speed camera~\cite{Juarez}. The recordings are performed under bright field at the recording speed of 125 frames per seconds (fps). To study the flow structures due to the nematode locomotion, fluorescent particles of diameter 2.0 $\mu$m are seeded in the fluid. These fluorescent particles are then recorded at 125 fps and tracked to resolve the velocity field induced by the nematode locomotion. In the granular media, the glass beads are tracked to reveal the particle transport pattern.

The experiment results show clear differences on the transport patterns emerged from these two media. As the nematode swims in the fluid media, four major recirculation regimes have been observed~\cite{XNS_visco}. When the nematode crawls through the granular media, the motility in monodisperse beads are quite similar to that of swimmming in fluids, and no significant voids in the media are observed. However, in the polydisperse beads, glass beads are seen transported perpendicular away from the nematode body and do not restore their original locations~\cite{Juarez}. The nematode leaves behind a band which does not contain glass beads.

The following videos are prepared for the web viewing and video gallery display.
\begin{enumerate}
  \item DFD2011\_Celegans\_ComplexMediaLo.avi
  \item DFD2011\_Celegans\_ComplexMediaHi.avi
\end{enumerate}

\bibliographystyle{amsplain}

\begin{thebibliography}{1}

\bibitem{Juarez}
G.~Juarez, K.~Lu, J.~Sznitman, and P.~E. Arratia, \emph{Motility of small
  nematodes in wet granular media}, EPL (Europhysics Letters) \textbf{92}
  (2010), 44002.

\bibitem{XNS_visco}
X.~N. Shen and P.~E. Arratia, \emph{Undulatory swimming in viscoelastic
  fluids}, Phys. Rev. Lett. \textbf{106} (2011), 208101.

\bibitem{JosuePF}
J.~Sznitman, X.~Shen, R.~Sznitman, and P.~E. Arratia, \emph{Propulsive force
  measurements and flow behavior of undulatory swimmers at low reynolds
  number},  \textbf{22} (2010), 121901.

\end{thebibliography}

\end{document}